\documentclass[12pt, draftclsnofoot, onecolumn]{IEEEtran}
\usepackage{amsmath}
\usepackage{amssymb}
\usepackage{amsthm}
\usepackage{cite}
\usepackage{color}
\usepackage{xcolor}
\usepackage{caption}
\usepackage{epsfig,latexsym}
\usepackage{float}
\usepackage{fancyhdr}
\usepackage{graphicx}
\usepackage{indentfirst}
\usepackage{mdwmath}
\usepackage{mdwtab}
\usepackage{subfigure}
\usepackage{setspace}
\usepackage{times}
\usepackage{url}
\usepackage{multirow}
\usepackage{algorithm}
\usepackage{algorithmic}

\newtheorem{remark}{Remark}
\newtheorem{theorem}{Theorem}

\newtheorem{lemma}{Lemma}

\newtheorem{corollary}{Corollary}

\newtheorem{proposition}{Proposition}

\begin{document}

\title{Performance Analysis of NOMA-RIS aided Integrated Navigation and Communication (INAC) Networks}
\author{Tianwei Hou,~\IEEEmembership{Member,~IEEE,}
        and Anna Li

%\thanks{ Manuscript received xx, 2023; revised xx 2022; accepted xx 2023. Date of publication ~~; date of current version~. }
\thanks{This work was supported by the National Natural Science Foundation for Young Scientists of China under Grant 62201028, and the Young Elite Scientists Sponsorship Program by CAST (Grant No.2022QNRC001). This paper is also supported by the Marie Skłodowska-Curie fellowship under Grant 101106428. This paper was submitted at the IEEE PIMRC 2023, Toronto, Canada, Sep. 2023~\cite{Hou2309:Integrated}. (Corresponding authors: Anna Li.)}
\thanks{T. Hou is with the School of Electronic and Information Engineering, Beijing Jiaotong University, Beijing 100044, China, and also with the Institute for Digital Communications, Friedrich-Alexander Universität Erlangen-Nürnberg (FAU), 91054 Erlangen, Germany (email: twhou@bjtu.edu.cn).}
\thanks{A. Li is with School of Electronic Engineering and Computer Science, Queen Mary University of London, London E1 4NS, U.K. (e-mail: anna.li@qmul.ac.uk).}
}

\maketitle

\begin{abstract}
Satellite communication constitutes a promising solution for the sixth generation (6G) wireless networks in terms of providing global communication services. In order to provide a cost-effective satellite network, we propose a novel medium-earth-orbit (MEO) satellite aided integrated-navigation-and-communication (INAC) network. To overcome the severe path loss of MEO satellites, we conceive a network for simultaneous serving navigation and communication for ground users by adopting the non-orthogonal multiple access (NOMA) technique and the reconfigurable intelligent surface technique. Based on the power allocation strategies, communication-oriented (CO-) and navigation-oriented (NO-) INAC scenarios are proposed. We first derive the closed-form expressions for the new channel statistics, outage probability and channel capacity of the INAC-user. For gleaning further insights, the diversity orders and navigation accuracy are evaluated for illustrating the performance of the INAC networks. According to our analysis, when RIS elements are sufficient, the proposed INAC network can perform better than conventional terrestrial communication networks in terms of channel capacity. Numerical results are provided for confirming that the NO-INAC and CO-INAC scenarios have superior performance for communication in the low signal-to-noise-ratio (SNR) regimes and high SNR regimes, respectively, which indicates a hybrid CO/NO-INAC network is preferable.
\end{abstract}

\begin{IEEEkeywords}
Integrated navigation and communication (INAC), NOMA, medium-earth-orbit (MEO) satellite, reconfigurable intelligent surface, 6G.
\end{IEEEkeywords}

\section{Introduction}
%6G needs global Service
In recent years, there has been an increase in the demand for sixth generation (6G) networks that offer global communication services and highly accurate positioning services~\cite{6G_1,6G_2}. Recently, the space-air-ground integrated networks (SAGIN) have been proposed by jointly combining unmanned aerial vehicle (UAV) communications, terrestrial communications and low-earth-orbit (LEO) satellite communications. In addition, satellites have been widely deployed for both global communications and global navigation~\cite{6G_satellite_mag1,6G_satellite_mag2,6G_satellite_mag3}.

In order to provide global wireless services, the low-earth-orbit (LEO) satellite communications stand as potential solutions for the 6G networks~\cite{Shanzi_satellite,satel_Comm_Mag}. Jung~{\em et al.} proposed a multiple-input-multiple-output (MIMO)-aided LEO satellite network, where shadowed Rician fading channels are utilized~\cite{satellite_shadow_2022}. Based on the fixed beamforming technique employed at the satellite, the small-scale channel gains of the MIMO channel can be degraded into a SISO channel with a larger antenna gain. Lin~{\em et al.} investigated a satellite physical layer security network, where the secrecy energy efficiency is optimized by utilizing the signal-to-leakage-plus-noise ratio (SLNR) metric~\cite{satellite_PLS_SLNR_R1_2}. However, due to the fact that the height of LEO satellites is lower than 500 km, the minimal required number of LEO satellites is even higher than 10000, which occupy a large number of LEO resources. Therefore, one natural question is: can we utilize medium-earth-orbit (MEO) satellites for providing wireless access to decrease the required number of satellites? Furthermore, Can we combine communication services into the navigation satellites for providing wireless access within the same bandwidth by exploiting the specific properties of the signal waveform?

%INAC is a good solution, and two main challenges
To enhance the cost efficiency of the LEO-satellite communication networks, integrated navigation and communication (INAC) networks performed at the MEO satellites stand as practical solutions~\cite{INAC_mag_2002_1,INAC_Mag_2018_2}. Theoretically, less than 100 MEO satellites are needed for global wireless communication services, where the number of MEO satellites is only 1\% to that of the LEO networks. In addition, communication and navigation functions can be integrated into a single hardware system, which can conserve space, weight and power consumption of MEO satellites.
Yin~{\em et al.} proposed an INAC network for D2D networks in~\cite{INAC_D2D}, and the fifth generation (5G) communications are coexisted with the global navigation satellite system (GNSS). Ma~{\em et al.} used a cyclic code shift keying (CCSK) for communication in an INAC network, where traditional navigation signal is used for pilot signal~\cite{CCSK_INAC}. Evans has proposed heterogeneous satellite communication networks in~\cite{LEO_MEO_satellite_person_com}, where both LEO and MEO satellites are deployed for personal communications. However, one inevitable problem is that the MEO satellites are located around 20000 km, hence resulting in severe path loss. Note that the performance of conventional amplify-then-forward (AF) relay and decode-then-forward (DF) relay in MEO satellite networks is still constrained by the deep fading channel of the satellite-relay links~\cite{Ssatellite_SINR_AF,satellite_SINR_relay}. This paper focuses on integrating navigation and communication functions into MEO satellites. Hence, no extra MEO satellites are required.
However, the INAC network supported by MEO satellites still faces two main challenges, which are: i) Since the orbit of MEO satellites is located at 20000 km, the path loss is greater than 180 dB, resulting in extremely low received signal-to-noise-ratio (SNR). Therefore, although MIMO technique is employed at the MEO satellites, the channel capacity of the MEO satellites still cannot support the requirement of communications. ii) The navigation signal has to be multi-casted to the ground users in all available time resource blocks (RBs), hence conventional time-divided multiple access (TDMA) may not be a practical solution.

%RIS-MEO can solve the first challenge,
Some new wireless technology, such as next generation multiple access (NGMA) and reconfigurable intelligent surfaces (RIS) has been proposed for 6G networks~\cite{RIS_survey_1,NGMA_JSAC_1}. The benefits of RIS-based techniques have gained considerable attention as a result of the emergence of new technologies in the 6G networks~\cite{RIS_survey_1,LIS_magazine_multi_scenarios,RIS_Mag_3}. Recently, RIS can be classified into two categories: signal-enhancement-based (SEB) and signal-cancellation-based (SCB) designs, where SCB design is employed for mitigating the interference in cellular networks~\cite{HOU_SCB_1,HOU_SSECB_2}. Tang~{\em et al.} proposed a SCB-design based physical layer security network~\cite{SCB_tang_PLS}, where the secrecy outage performance is optimized by minimizing the summation of the reflected signals and desired signal of the Eve. The SEB design, in contrast, is primarily concerned with enhancing the desired signal power level through the appropriate adjustment of reflection angles and amplitude coefficients of the RIS elements, resulting in ``square-law'' improvement. Since the received SINR of commercial 5G networks is between 3 dB to 23 dB by using advanced MIMO techniques, the SEB design of RIS has to be applied in a more valuable scenario~\cite{SINR_5G}. For example, the RIS technique is needed and expected for those communication scenarios with SINR lower than -10 dB, which can be naturally composed with satellite communications~\cite{RIS_satellite_1}. Guo~{\em et al.} proposed a multi-tier satellite and unmanned aerial vehicle (UAV) network in~\cite{RIS_satellite_UAV}, where RIS is deployed for improving the signal power level. Hou~{\em et al.} proposed a stochastic geometry aided RIS-MIMO network~\cite{Hou_RIS_MIMO_global_algrithm}, where multiple users are served equally.

In order to simultaneously provide navigation and communication services, navigation and communication signals must be composed into all available time RBs. On the one hand, as a promising technique for opportunistically serving multiple users with different functions at the same time/frequency/code RBs, non-orthogonal multiple access (NOMA) has been proposed~\cite{NOMA_5G_beyond_Liu,NOMA_mag_Ding,Islam_NOMA_survey,NOMA_large_heter}. On the other hand, the NOMA technique is more proper for the energy constraint scenarios due to its higher energy efficiency, e.g., SAGIN~\cite{NOMA_SAGIN}. Hence, motivated by the aforementioned requirements, the NOMA technique is employed to simultaneously transmit both communication and navigation signals. Yue~{\em et al.} evaluated the outage performance of NOMA-aided satellite networks in~\cite{Xinwei_satel_NOMA}, where the shadowed-Rician fading channels are utilized. Lin~{\em et al.} proposed a satellite-terrestrial network, where the NOMA technique is employed for simultaneously serving more ground users~\cite{NOMA_SAGIN_JSTSP_R1_1}. Furthermore, the rate-splitting multiple access technique is used in SAGIN to enhance the performance of internet-of-things (IoT)~\cite{RSMA_SAGIN_R1_3}.
Yan~{\em et al.} proposed a relay-aided satellite network in~\cite{satellite_NOMA_relay}. It is also indicated that the active beamforming gain can be simplified to the transmitting antenna gain. Zhu~{\em et al.} proposed a NOMA-based integrated terrestrial-satellite networks~\cite{NOMA_terre_satel}, and the satellites acted as the supplementary of the conventional terrestrial communication networks. Yue~{\em et al.} proposed a NOMA-aided RIS network over Rician fading channels, where imperfect channel state information has been considered \cite{Xinwei_STAR_RIS_NOMA}. Guo~{\em et al.} proposed a NOMA-aided RIS for the non-terrestrial vehicle networks, where the exact and asymptotic expressions of outage probability are derived~\cite{NOMA_RIS_SAGIN_LIURUI1}. Singh~{\em et al.} proposed a NOMA-aided two-way relay integrated-satellite-terrestrial network, where cache-free and cache-aided scenarios are investigated~\cite{NOMA_satellite_R2_1}. Furthermore, a NOMA-aided integrated-satellite-terrestrial network for physical layer security was proposed in~\cite{NOMA_Satel_relay_PLS_R2_2}, where the closed-form expressions of secrecy outage probability (SOP) are analysed by adopting shadowed Rician fading channels.

%To further enhance both the SE and EE of the DL, NOMA and RIS techniques were integrated in~\cite{DING_RIS_NOMA_letter}. The RISs can be deployed for enhancing the power level of the cell-edge users, where the cell-center users treat the reflected signal as interference~\cite{DING_RIS_NOMA_letter}. Both continuous and discrete phase shifters were used in a RIS-aided MISO NOMA network~\cite{yuanwei_NOMA_RIS}. Naturally, the BS-user link plays a key role~\cite{MISO_with_directlink}. A RIS-aided NOMA network was also investigated in~\cite{NOMA_RIS_Fu}, whilst the BS-user link and the BS-RIS link, as well as the RIS-user link were assumed to experience Rayleigh fading. The associated bit error ratio (BER) was evaluated in the case of Rayleigh fading in~\cite{LIS_perform_Anal}.
%However, both the BS and RISs are part of the infrastructure, and the RISs are typically positioned for exploiting the line-of-sight (LoS) path with respect to the fixed BS in NG networks for increasing the received signal power. Hence, the impact of fading environments on RIS networks has also attracted attention~\cite{RIS_NOMA_Rice}.
%A fairness-oriented algorithm was proposed in a RIS-aided NOMA network~\cite{RIS_NOMA_Rice}, where Rician fading channels were used for modelling the channel gains. Note that when the Nakagami and Rice fading parameter obey the following constraint $m=\frac{(K+1)^2}{2K+1}$, these fading channels are identical~\cite[eq. (3.38)]{wireless_communication_goldsmith}.

%Motivation
Among the papers mentioned above, RIS-assisted or NOMA-assisted satellite communication networks have been studied extensively, whereas there have been limited investigations into NOMA-RIS-aided INAC networks. To provide a cost-effective satellite network, the NOMA-RIS-aided INAC network poses three additional challenges: i) In order to provide a cost-efficient satellite network, communication function and navigation function are expected to be integrated into MEO satellites; ii) Since the navigation function has higher priority for the MEO satellites, NOMA assisted navigation and communication composition policy is still unknown; iii) The signal transmitted by the MEO satellites cannot provide enough power level to satisfy the minimum requirement for communications, where RIS technique provides a potential solution for the proposed networks. This article aims at tackling the aforementioned issues, by integrating navigation and communication into MEO satellites, proposing two potential power allocation scenarios, namely communication-oriented (CO) and navigation-oriented (NO) scenarios, for intelligently investigating the effect of NOMA-RIS-aided INAC network.

%Inspired by the aforementioned benefits of NOMA and RIS techniques, we explore the network's performance enhanced by the intrinsic integration of the power-domain NOMA

The following is a brief summary of our main theoretical contributions with regard to the MEO satellite set-up in this article:
\begin{itemize}
  \item  We develop a novel NOMA-RIS-aided INAC network, where only 40 MEO satellites are needed for global communication services. The NOMA technique is utilized for simultaneous transmitting the desired communication and navigation signals, where CO-INAC and NO-INAC scenarios are proposed for better illustrating the appropriate applications. The proposed technique is capable of conserving the required space, weight and power consumption of satellites, which has considerable research and engineering prospectives.
  \item The RIS is deployed on the ground to improve the small-scale channel gains between MEO satellites and ground users for improving the channel capacity of the INAC networks. This paper evaluates the impact of line-of-sight (LoS) links and the number of RIS elements on reflected satellite-RIS-user links. Furthermore, the proposed design is evaluated with respect to outage probability (OP), channel capacity, and navigation accuracy to determine its impact on attainable performance.
  \item We then derive the OP and the channel capacity for the proposed NOMA-RIS-aided INAC networks. Both accurate and approximate expressions of the OP and channel capacity of the CO-INAC and NO-INAC scenarios are derived. Moreover, the diversity orders are calculated based on the OP, and these can be enhanced by varying the number of RIS elements or the fading parameter. Our analytical results indicate that the central limit theorem (CLT) can provide accurate results in the low-SNR regime.
  \item The simulation results confirm our analysis, demonstrating that: 1) the proposed NOMA-RIS-aided INAC networks can provide excellent performance when the number of RIS elements is high enough; 2) the hybrid CO/NO-INAC may be a good solution because that the CO-INAC scenario and NO-INAC scenario can provide better channel capacity in the high-SNR and low-SNR regimes, respectively; 3) the navigation accuracy of the proposed NO/CO-INAC scenarios are identical in the high-SNR regimes, which indicates that the increased number of RIS elements has nearly no impact on navigation accuracy.
\end{itemize}

\subsection{Organization and Notations}

The remainder of this article is organized as follows. Section \uppercase\expandafter{\romannumeral2} discusses the model of NOMA-RIS-aided INAC networks. Our analytical results of the CO-INAC and NO-INAC scenarios are presented in Section \uppercase\expandafter{\romannumeral3} and Section \uppercase\expandafter{\romannumeral4}, respectively. Section \uppercase\expandafter{\romannumeral5} provides the numerical results of our proposed INAC networks. In Section \uppercase\expandafter{\romannumeral6}, we conclude our article. $\sim$ stands for ``distributed as''. $\mathbb{P}(\cdot)$ and $\mathbb{E}(\cdot)$ denote the probability and expectation, respectively. %Table~\ref{TABLE OF NOTATINS} lists the main notations used in this article.

%\begin{table}
%\caption{\\TABLE OF NOTATIONS}
%\centering
%\begin{tabular}{|l|r|}
%\hline
%$R_W$ &  The target rate of user $W$. \\
%\hline
%$m_1$ &  Fading parameter between the BS and RISs.\\
%\hline
%$m_W$ &  Fading parameter between the RISs and user $W$.\\
%\hline
%$W$ & The number of users.\\
%\hline
%$N$ & The number of RISs. \\
%\hline
%$\rm \bf h$  & The channel vector between the BS and RISs. \\
%\hline
%${\rm \bf g}_W$ & The channel vector between the RISs and user $W$. \\
%\hline
%${r}_W$ & The channel gain between the BSs and user $W$.\\
%\hline
%$\rm \bf \Phi$ & The effective matrix of RISs.\\
%\hline
%\end{tabular}
%\label{TABLE OF NOTATINS}
%\end{table}

\section{System Model}

\begin{figure}[t!]
\centering
\includegraphics[width =3in]{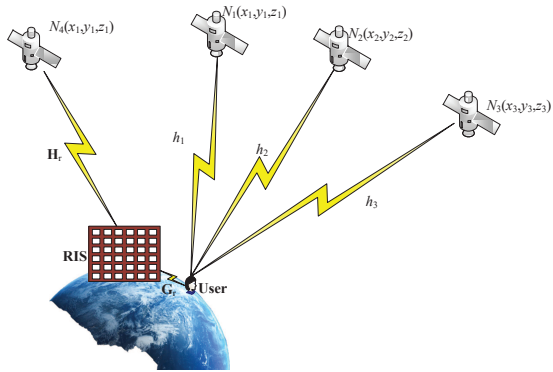}
\caption{Illustration of NOMA-RIS-aided INAC networks.}
\label{system_model}
\end{figure}

In this article, in order to better illustrate the performance gain obtained by the proposed INAC networks, we consider a NOMA-RIS-aided INAC network, where each satellite equipped with $M$ transmitting antennas (TAs) is communicating with two ground users, each equipped with a single receiving antenna (RA). The number of RIS elements $L>1$ is deployed near the ground users at the appropriate location. We then comprehensively analyse the performance of MEO satellite-aided INAC networks. Fig.~\ref{system_model} illustrates the proposed NOMA-RIS-aided INAC model.

\subsection{Signal Model based on NOMA}

In next generation networks, the spectral efficiency and energy efficiency are required to outperform that of the conventional orthogonal multiple access (OMA) networks, where the NOMA technique is expected as a potential solution for provisioning unprecedented massive user access. In this article, since both low-earth-orbit and high-earth-orbit satellites cannot provide enough time dilution of precision (TDoP) and position dilution of precision (PDoP) for navigation, we only consider that MEO satellites provide both navigation and communication services to the ground users. In downlink transmission, due to the fact that navigation signals are required for all ground users, the navigation signals are defined as multi-cast signals. On the contrary, each user requires different communication signals, which can be defined as the uni-cast signals in the proposed INAC networks. Therefore, the composed signal can be written as:
\begin{equation}\label{information bearing}
{\bf{s}} = {\alpha _M}{{\bf{s}}_M} + {\alpha _U}{{\bf{s}}_U},
\end{equation}
where $\mathbf{s}_M$ and $\mathbf{s}_U$ denote the signals intended for multi-cast and uni-cast, respectively, with $\alpha _M$ and $\alpha _U$ representing the power allocation factors of multi-cast and uni-cast signals with $\alpha _M^2+ \alpha _U^2 =1$, respectively.

Based on the NOMA protocol, NOMA-aided INAC networks are typically classified into two categories~\cite{RIS_two_scenarios}, i.e., CO-INAC and NO-INAC. When navigation stands as the first priority, less power should be allocated to navigation based on the superposition coding (SC) and successive interference cancellation (SIC). By contrast, when communication constitutes the first priority, more power should be allocated to navigation for improving the performance of communication.

\subsection{Channel Model}
A fundamental model is provided to illustrate the performance of the INAC networks affected by RISs. In the proposed INAC networks, three navigation satellites $n=1,2,3$ and one INAC satellite are deployed in space for providing both navigation and communication services, and the satellites are distributed on the surface of a sphere at the MEO. In this article, we assume that the channel state information (CSI) is well known at the RIS~\cite{ZhangRui_MISO_beams_discrete_2,MIMO_channel_estimation}.

\begin{figure}[t!]
\centering
\includegraphics[width =2.5in]{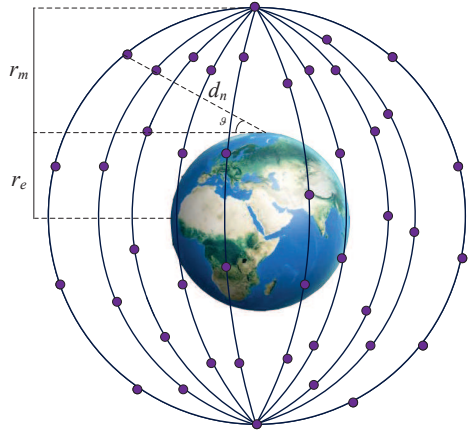}
\caption{Illustration of satellites and earth.}
\label{earth_model}
\end{figure}

As shown in Fig.~\ref{earth_model}, the radius of the navigation satellites located at MEO can be given by $r_e+r_m$, where $r_e$ and $r_m$ denote the radius of Earth and the height of satellites. The elevation angle between the satellite and ground user is set to $\vartheta $, and hence by utilizing the law of Cosines, the distance between the satellite and ground user can be given by:
\begin{equation}\label{distance between satel to user}
d_n = \sqrt {r_e^2{{\sin }^2}\vartheta  + r_m^2 + 2{r_e}{r_m}}  - {r_e}\sin \vartheta.
\end{equation}

Since the RIS array is usually located on the ground or building surfaces, the large-scale fading channel of satellite-user and satellite-RIS link can be treated as identical, hence the large-scale fading channel of satellite-user and satellite-RIS links is defined as:
\begin{equation}\label{large scale fading channel,satellite to user}
l\left( {{d_n}} \right) = {G_T}{(\frac{c}{{4\pi {f_c}}})^2}d_n^{ - {\alpha _1}},
\end{equation}
where ${G_T}$ represents the gain of TAs. $c $ and $f_c$ represent the speed of light and the carrier frequency, respectively. $\alpha _1$ denotes the path loss exponent between satellite and user/RIS.

Similarly, the large-scale fading channel of the RIS-user link is defined as:
\begin{equation}\label{large scale fading channel,RIS to user}
l\left( {{d_{{\rm{RU}}}}} \right) = {(\frac{c}{{4\pi {f_c}}})^2}d_{{\rm{RU}}}^{^{ - {\alpha _2}}},
\end{equation}
where $d_{{\rm{RU}}}$ represents the distance between RIS and ground user. ${\alpha_2}$ denotes the path loss exponent of RIS-user link. Note that the MEO satellite mainly considers fixed-beam strategy at the satellites, the beam is directed to the projection point of the satellite. In this article, the antenna gain ${G_T}$ can be set to 32 dBi in the main lobe.

We then turn our focus on the small-scale fading channels. Naturally, in navigation channels, since the active beamforming at satellites is fixed, the small-scale matrix can be degraded to a single variable with larger TA gains. In practice, many mathematical models have been developed to describe the satellite-terrestrial channel, such as Rician, Loo, Barts–Stutzman, and Karasawa models. Therefore, since no objects can block the links between navigation satellites $n$ and ground users, hence the small-scale fading channels ${{h_{n}}}$ for $n=1,2,3$ are considered as Rician fading channels as follows:
\begin{equation}\label{Rice channel gain,sateli to user}
h_{n} = \sqrt {\frac{ \mathcal{K}_n }{{\mathcal{K}_n + 1}}} h_{n}^{{\rm{LoS}}} + \sqrt {\frac{1}{{\mathcal{K}_n + 1}}} h_{n}^{{\rm{NLoS}}},
\end{equation}
where $\mathcal{K}_n$ represents the Rician fading parameter of the satellite-user links. The LoS and Non-LoS (NLoS) components are defined by $h_{n}^{{\rm{LoS}}} $ and $h_{n}^{{\rm{NLoS}}} $, respectively.

In this article, $L$ RIS elements provide signal enhancement services to ground user for improving the performance of communications. Similarly, the small scale fading channel of satellite-RIS link also follows Rician fading channels, which is given by
\begin{equation}\label{small scale fading channel,satellite to RIS}
{{\bf{H}}_{\rm{r}}} = \sqrt {\frac{{{{\cal K}_r}}}{{{{\cal K}_r} + 1}}} {\bf{\bar H}}_{\rm{r}}^{{\rm{LoS}}} + \sqrt {\frac{1}{{{{\cal K}_r} + 1}}} {\bf{\bar H}}_{\rm{r}}^{{\rm{NLoS}}},
\end{equation}
where $\mathcal{K}_{r}$ denotes the Rician factor of the satellite-RIS links. ${{\bf{H}}_{\rm{r}}}$, ${\bf{\bar H}}_{\rm{r}}^{{\rm{LoS}}}$ and ${\bf{\bar H}}_{\rm{r}}^{{\rm{NLoS}}}$ are $\left( {L \times 1} \right)$-element vectors. ${\bf{\bar H}}_{\rm{r}}^{{\rm{LoS}}}$ and ${\bf{\bar H}}_{\rm{r}}^{{\rm{NLoS}}}$ denote LoS and NLoS components containing elements $h_{{\rm{r}},l}^{{\rm{LoS}}}$ and $h_{{\rm{r}},l}^{{\rm{NLoS}}}$, respectively.
%More specifically, each element in the channel matrix  ${{\bf{H}}_{\rm{r}}}$ of satellite-RIS links contains LoS and NLoS components as follows:
%\begin{equation}\label{Rice channel gain,satel to RIS}
%{h_{{\rm{r}},l}} = \sqrt {\frac{{{{\cal K}_r}}}{{{{\cal K}_r} + 1}}} h_{{\rm{r}},l}^{{\rm{LoS}}} + \sqrt {\frac{1}{{{{\cal K}_r} + 1}}} h_{{\rm{r}},l}^{{\rm{NLoS}}}.
%\end{equation}

Furthermore, the small-scale channel vector between RIS and ground user can be given by:
\begin{equation}\label{small scale fading channel, RIS to user}
{{\bf{G}}_{\rm{r}}} = \sqrt {\frac{{{{\cal K}_g}}}{{{{\cal K}_g} + 1}}} {\bf{\bar G}}_{\rm{r}}^{{\rm{LoS}}} + \sqrt {\frac{1}{{{{\cal K}_g} + 1}}} {\bf{\bar G}}_{\rm{r}}^{{\rm{NLoS}}},
\end{equation}
where ${{\bf{G}}_{\rm{r}}}$ denotes a $\left( {1 \times L} \right)$-element vector containing elements $g_{{\rm{g}},l}^{{\rm{LoS}}}$ and $g_{{\rm{g}},l}^{{\rm{NLoS}}}$, respectively.

Without loss of generality, we focus our attention on the INAC-user, and the signal received by the INAC-user from the satellite through RISs is given by~\cite{RIS_two_scenarios}
\begin{equation}\label{received user signal}
y = {{\bf{G}}_{\rm{r}}}{\bf{\Phi }}{{\bf{H}}_{\rm{r}}}l\left( d \right)l\left( {{d_{{\rm{RU}}}}} \right)p {g_{{\rm{sp}}}}{\bf{s}} + {N_0},
\end{equation}
where ${p}$ denotes the transmit power of satellite, ${g_{{\rm{sp}}}}$ represents the spread spectrum gain, ${\rm \bf \Phi}  \buildrel \Delta \over = {\rm{diag}}\left[ {{\beta_1} {\phi _1}, {\beta_2}{\phi _2}, \cdots,{\beta_{L}} {\phi _{L}}} \right]$ denotes reflection matrix of RIS. $\beta_l  \in \left( {0,1} \right]$ represents the amplitude coefficient of RISs. In order to obtain tractable analytical results, we set $\beta_1,\ldots,\beta_L=1$ in~\eqref{received user signal} for simplicity. ${\phi _l} = \exp (j{\theta _l}), \forall l = 1,2 \cdots ,L$, and ${\theta _l} \in \left[ {0,2\pi } \right)$ are the phase shift of RIS element $l$. The additive white Gaussian noise (AWGN) is represented by $N_0$ denotes with variance ${\sigma ^2}$. Since atomic clocks are deployed at the MEO navigation satellites, which can provide high accurate carrier frequency, the Doppler frequency shift of MEO satellites can be calculated at mHz level. Therefore, we simply ignore the Doppler frequency shift in~\eqref{received user signal} by high-order Doppler detection algorithm~\cite{Higher_order_doppler}.

%
%\begin{table}
%\caption{\\ TABLE OF SYMBOLS}
%\centering
%\begin{tabular}{|l|r|}
%\hline
%$\left| {{{{\rm{\tilde h}}}_{W,l}}} \right|^2$ &  The worst-case distribution of user $W$. \\
%\hline
%$\left| {{{{\rm{\tilde h}}}_{W,u}}} \right|^2$ &  The best-case distribution of user $W$.\\
%\hline
%${P_{W,u}}$ & The worst-case of the OP expression of user $W$.\\
%\hline
%${P_{W,l}}$ & The best-case of the OP expression of user $W$.\\
%\hline
%${{\overline{P}}_{W,u}}$ & The approximated worst-case OP of user $W$. \\
%\hline
%${{\overline P}_{W,l}}$  & The approximated best-case OP of user $W$.\\
%\hline
%$d_{W,u}$ & The diversity order of the worst-case of user $W$.\\
%\hline
%$d_{W,l}$ & The diversity order of the best-case of user $W$.\\
%\hline
%$R_{W,l}$ & The worst-case of the ergodic rate of user $W$.\\
%\hline
%$R_{W,u}$ & The best-case of the ergodic rate of user $W$. \\
%\hline
%\end{tabular}
%\label{TABLE OF math symbols}
%\end{table}

\section{Performance Analysis of the CO-INAC Networks}

This section begins with the passive beamforming design for the RISs. Since the data requirement of navigation is much lower than communications, the passive beamforming at RIS is simply designed for ensuring the INAC-user has the best channel gains by designing the phase shifts as follows:
\begin{equation}\label{define of the passive beamforming}
\begin{aligned}
& max\left| {{{\bf{G}}_{\rm{r}}}{\bf{\Phi }}{{\bf{H}}_{\rm{r}}}} \right|\\
& subject\;to\;{\beta _{1}} \cdots {\beta _{L}} = 1\\
& {\theta _{1}} \cdots {\theta _{L}} \in \left[ {0,2\pi } \right).
\end{aligned}
\end{equation}
Thus by utilizing the signal alignment technique, and similar to~\cite{NOMA_RIS_JSAC_HOU}, our objective can be achieved by phase-shifting the signals received at the RISs, which is capable of significantly improving the received power. In order to design the phase shift, we first define a channel vector as follows:
\begin{equation}\label{channel matrix }
{\bf{\tilde h}} = \left[ {\begin{array}{*{20}{c}}
{ {g_{{\rm{g}},1}}    {h_{{\rm{r}},1}} }& \cdots &{ {g_{{\rm{g}},l}} {h_{{\rm{r}},l}}  }
\end{array}} \right].
\end{equation}

Thus, the phase shifts of the RISs can be further transformed into
\begin{equation}\label{RIS phase shift design}
{\bf{\Phi }}{\rm{ = }}  {\theta _{\rm{d}}} - \arg ({\bf{\tilde h}}),
\end{equation}
where $\arg( \cdot )$ denotes the angle of the element. ${\theta _{\rm{d}}}$ represents the objective phase, which can be any value.

By doing so, the INAC-user has the best channel gain, and the reflected satellite-RIS-user (SRU) signals are co-phased at the INAC-user, which is given by:
\begin{equation}\label{maximum achievable gain}
\max {{\bf{G}}_{\rm{r}}}{\bf{\Phi }}{{\bf{H}}_{\rm{r}}} = \sum\limits_{l = 1}^L {\left| {{\beta _l}{h_{{\rm{r}},l}}{g_{{\rm{r}},l}}} \right|} .
\end{equation}

\subsection{New Channel Statistics}
The purpose of this subsection is to derive new channel statistics for the proposed NOMA-RIS-aided INAC networks, which will serve as a basis for evaluating the OPs.

\begin{lemma}\label{lemma1:new state of effective channel gain}
When the elements of the channel vectors are independently and identically distributed (i.i.d.), The PDF of the effective channel gain of the INAC-user can be formulated as
\begin{equation}\label{New Gamma distribution PDF}
\begin{aligned}
&  {f_{{{\left| {\tilde h} \right|}^2}}}\left( x \right) = \frac{1}{{2\sqrt {2\pi {v_3}x} }} \\
& \times \left( {\exp \left( { - \frac{{{{\left( {\sqrt x  + {m_3}} \right)}^2}}}{{2{v_3}}}} \right) + \exp \left( { - \frac{{{{\left( {\sqrt x  - {m_3}} \right)}^2}}}{{2{v_3}}}} \right)} \right),
\end{aligned}
\end{equation}
where ${m_3} =\mathbb{E} \left\{ {\tilde h} \right\} = {\beta _l}L{m_1}{m_2}$ and ${v_3} = {\beta _l}L\left( {m_1^2{v_2} + m_2^2{v_1} + {v_1}{v_2}} \right)$. ${m_1} = \sqrt {\frac{\pi }{{4\left( {1 + {{\cal K}_{r,l}}} \right)}}} {}_1{F_1}\left( { - \frac{1}{2},1;{{\cal K}_{r,l}}} \right)$, ${m_2} = \sqrt {\frac{\pi }{{4\left( {1 + {{\cal K}_{g,l}}} \right)}}} {}_1{F_1}\left( { - \frac{1}{2},1;{{\cal K}_{g,l}}} \right)$. ${v_1} = 1 - \frac{\pi }{{4\left( {1 + {{\cal K}_{r,l}}} \right)}}{}_1{F_1}{\left( { - \frac{1}{2},1;{{\cal K}_{r,l}}} \right)^2}$, ${v_2} = 1 - \frac{\pi }{{4\left( {1 + {{\cal K}_{g,l}}} \right)}}{}_1{F_1}{\left( { - \frac{1}{2},1;{{\cal K}_{g,l}}} \right)^2}$.
The Cumulative Density Function (CDF) can be expressed by
\begin{equation}\label{New Gamma distribution CDF}
{F_{{{\left| {\tilde h} \right|}^2}}}\left( x \right) = \frac{1}{2}\left( {{\rm{erf}}\left( {\frac{{\sqrt x  + {m_3}}}{{\sqrt {2{v_3}} }}} \right) - {\rm{erf}}\left( {\frac{{-\sqrt x +  {m_3}}}{{\sqrt {2{v_3}} }}} \right)} \right),
\end{equation}
where ${\rm{erf}}\left(  \cdot  \right)$ represents the error function.
\begin{proof}
Please refer to Appendix A.
\end{proof}
\end{lemma}

\begin{remark}\label{Channel gain remark1}
 The results of~\eqref{New Gamma distribution PDF} demonstrate that the small-scale channel gain of the NOMA-RIS-aided INAC networks is much higher than that of conventional satellite communications, which dramatically improves the spectral efficiency. In addition, when the number of RIS elements is high enough, the spread-encoding/decoding techniques are not necessary for the proposed NOMA-RIS-aided INAC networks.
\end{remark}

\subsection{SINR of the CO-INAC scenario}
In the CO-INAC scenario, more power is allocated to the multi-cast signal. Therefore, the INAC-user first detects the multi-cast signal by using the SIC technique with the following SINR:
\begin{equation}\label{SINR_CO_Navigation}
SIN{R_{{\rm{CO,N}}}} = \frac{{\alpha _M^2{{\left| {\tilde h} \right|}^2}l\left( d \right)l\left( {{d_{{\rm{RU}}}}} \right)p{g_{{\rm{sp}}}}}}{{\alpha _U^2{{\left| {\tilde h} \right|}^2}l\left( d \right)l\left( {{d_{{\rm{RU}}}}} \right)p{g_{{\rm{sp}}}} + {\rho ^2}}}.
\end{equation}

If the multi-cast signal can be successively detected and deleted by the SIC technique, the INAC-user can then detect the uni-cast signal with the following SINR:
\begin{equation}\label{SINR_CO_communication}
SIN{R_{{\rm{CO,C}}}} = \frac{{\alpha _U^2{{\left| {\tilde h} \right|}^2}l\left( d \right)l\left( {{d_{{\rm{RU}}}}} \right)p{g_{{\rm{sp}}}}}}{{{\rho ^2}}}.
\end{equation}

\subsection{Outage Probability}
In this article, the OP is defined as the probability that the information rate attained of uni-cast or multi-cast is lower than the required target rate. In the CO-INAC scenario, the INAC-user needs to first detect the multi-cast signal with the OP:
\begin{equation}\label{OP_CO_INAC_detect NO_signal}
{P_{{\rm{CO}},{\rm{N}}}} =\mathbb{P} \left( {{{\log }_2}(1 + SIN{R_{{\rm{CO,N}}}}) < {R_M}} \right),
\end{equation}
where ${{R_M}}$ represents the target rate of multi-cast. By utilizing SIC, when the INAC-user successfully detects the multi-cast signal, the uni-cast signal can be then detected with the OP:
\begin{equation}\label{OP_CO_INAC_detect CO_signal}
\begin{array}{l}
{P_{{\rm{CO,C}}}} =\mathbb{P}  \left( {{{\log }_2}(1 + SIN{R_{{\rm{CO,N}}}}) < {R_M}} \right)\\
 + \left( {{{\log }_2}(1 + SIN{R_{{\rm{CO,N}}}}) > {R_M},{{\log }_2}(1 + SIN{R_{{\rm{CO,C}}}}) < {R_U}} \right),
\end{array}
\end{equation}
where $R_U$ represents the target rate of uni-cast.

The next step is to calculate the OP of the CO-INAC scenario based on the Theorems $1$ and $2$.
\begin{theorem}\label{Theorem1:Outage CO-INAC decoding communication closed form by error func}
\emph{In the CO-INAC scenario, assuming that $ \alpha _M^2 - \alpha _U^2{\varepsilon _M} > 0$, the closed-form OP expression for decoding the multi-cast signal can be expressed as}
\begin{equation}\label{outage analytical results CO-INAC decode multicast in theorem1}
\begin{aligned}
{P_{{\rm{CO,N}}}} =  \frac{1}{2}\left( {{\rm{erf}}\left( {\frac{{\sqrt {{\omega _{\rm{CO,N}}}}  + {m_3}}}{{\sqrt {2{v_3}} }}} \right) - {\rm{erf}}\left( {\frac{{m_3}-{\sqrt {{\omega _{\rm{CO,N}}}}   }}{{\sqrt {2{v_3}} }}} \right)} \right),
\end{aligned}
\end{equation}
\emph{where ${\omega _{{\rm{CO,N}}}} = \frac{{{\varepsilon _M}{\rho ^2}}}{{\left( { \alpha _M^2 - \alpha _U^2{\varepsilon _M}} \right)\gamma }}$, $\gamma  = l\left( d \right)l\left( {{d_{{\rm{RU}}}}} \right)p{g_{{\rm{sp}}}}$, ${\varepsilon _M} = {2^{{R_M}}} - 1$, ${\varepsilon _U} = {2^{{R_U}}} - 1$.}
\begin{proof}
Please refer to Appendix B.
\end{proof}
\end{theorem}

\begin{theorem}\label{Theorem2:Outage CO-INAC  decoding navigation  closed form by error func}
\emph{In the CO-INAC scenario, assuming that $  \alpha _M^2 - \alpha _U^2{\varepsilon _M} > 0$, the closed-form OP expression for decoding the uni-cast signal can be expressed as}
\begin{equation}\label{outage analytical results CO-INAC decode multicast in theorem2}
\begin{aligned}
{P_{{\rm{CO,C}}}} =  \frac{1}{2}\left( {{\rm{erf}}\left( {\frac{  {m_3}+ {\sqrt {{\omega _{\rm{CO,C}}}} }}{{\sqrt {2{v_3}} }}} \right) - {\rm{erf}}\left( {\frac{{m_3}-{\sqrt {{\omega _{\rm{CO,C}}}}  }}{{\sqrt {2{v_3}} }}} \right)} \right),
\end{aligned}
\end{equation}
\emph{where ${\omega _{{\rm{CO,C}}}} = \max \left\{ {\frac{{{\varepsilon _M}{\rho ^2}}}{{\left( { \alpha _M^2 - \alpha _U^2{\varepsilon _M} } \right)\gamma }},\frac{{{\varepsilon _U}{\rho ^2}}}{{\alpha _U^2\gamma }}} \right\}$.}
\begin{proof}
Similar to Appendix B, the proof is complete.
\end{proof}
\end{theorem}

However, the error function makes it difficult to obtain direct engineering insights from ~\eqref{outage analytical results CO-INAC decode multicast in theorem1} and~\eqref{outage analytical results CO-INAC decode multicast in theorem2}. Consequently, the approximate behavior of the high-SNR regime is analyzed in order to gain further insight.

\begin{corollary}\label{corollary1:Outage CO_INAC_navigation}
\emph{Assuming that $ \alpha _M^2 - \alpha _U^2{\varepsilon _M} > 0$, the OP of the INAC-user for decoding the multi-cast signal in the CO-INAC scenarios can be approximated by}
\begin{equation}\label{Corro1:CO_INAC_decoding navigation}
\begin{aligned}
{{\bar P}_{{\rm{CO,N}}}} =  & \frac{4}{{\sqrt \pi  }}\sum\limits_{n = 0}^\infty  {\frac{{{{\left( { - 1} \right)}^n}}}{{n!\left( {2n + 1} \right){{\left( {2{v_3}} \right)}^{\frac{{2n + 1}}{2}}}}}} \\
&\times \sum\limits_{k = 1,3...2n + 1}^{} {\left( {\begin{array}{*{20}{c}}
{2n + 1}\\
k
\end{array}} \right)} m_3^{2n + 1 - k}\omega _{{\rm{CO,N}}}^{\frac{k}{2}},
\end{aligned}
\end{equation}
\begin{proof}
Please refer to Appendix C.
\end{proof}
\end{corollary}

\begin{corollary}\label{corollary2:Outage CO_INAC for decoding navigation}
\emph{Assuming that $ \alpha _M^2 - \alpha _U^2{\varepsilon _M} > 0$, the OP of the INAC-user for decoding the uni-cast signal in the CO-INAC scenarios can be approximated by}
\begin{equation}\label{Corro2:CO_INAC_decoding communication}
\begin{aligned}
{{\bar P}_{{\rm{CO,C}}}} =  & \frac{4}{{\sqrt \pi  }}\sum\limits_{n = 0}^\infty  {\frac{{{{\left( { - 1} \right)}^n}}}{{n!\left( {2n + 1} \right){{\left( {2{v_3}} \right)}^{\frac{{2n + 1}}{2}}}}}} \\
&\times \sum\limits_{k = 1,3...2n + 1}^{} {\left( {\begin{array}{*{20}{c}}
{2n + 1}\\
k
\end{array}} \right)} m_3^{2n + 1 - k}\omega _{{\rm{CO,C}}}^{\frac{k}{2}},
\end{aligned}
\end{equation}
\begin{proof}
Similar to Appendix C, the proof is complete.
\end{proof}
\end{corollary}

\begin{remark}\label{remark2:Asymptotic condition}
Since the asymptotic results are derived based on the Taylor expansion of the error function in~\eqref{Appendix C Lower incomplete gamma expansion}, $z$ should be smaller than 1, otherwise, the asymptotic result does not exist.
\end{remark}

It is possible to obtain the diversity orders of the INAC-user for the following propositions in order to evaluate the slope of the OP.
\begin{proposition}\label{proposition1: diversity order CO_INAC}
\emph{Based on \textbf{Corollary~\ref{corollary1:Outage CO_INAC_navigation}} and~\textbf{Corollary~\ref{corollary2:Outage CO_INAC for decoding navigation}}, the diversity order of the INAC-user in the CO-INAC scenarios is given by}
\begin{equation}\label{diversity order of w}
{d_{{\rm{CO,C}}}} =  {d_{{\rm{CO,N}}}} = - \mathop {\lim }\limits_{\frac{p}{{{\sigma ^2}}} \to \infty } \frac{{\log {{\bar P}_{{\rm{CO,C}}}}}}{{\log \frac{p}{{{\sigma ^2}}}}} \approx m_3.
\end{equation}
\end{proposition}

\begin{remark}\label{remark2:impact of fading environment on diversity order}
Based on the results of \textbf{ Proposition~\ref{proposition1: diversity order CO_INAC}}, in the proposed NOMA-RIS-aided INAC networks, the diversity order is mainly affected by the fading parameters of satellite-RIS and RIS-user links, as well as the number of RIS elements.
\end{remark}

\begin{remark}\label{remark3:strong los impact}
Assuming that the elevation angle $\vartheta $ is small enough, which indicates a strong LoS link exists between the satellite and RISs ${{{\cal K}_r}} \to \infty $, and the number of RIS elements is high enough, the diversity orders of the INAC-user is $Lm_2$.
\end{remark}

\subsection{Channel Capacity}

In this article, since the channel capacity of the INAC-user is complicated, we simply use channel hardening to derive tractable rate performance. Based on the insights from~\cite{chennel_corralt}, when the elements of ${h}$ are i.i.d. with mean $m_1$ and $m_2$, then the expectation of ${h}$ can be given by~\cite{Channel_expectation_hardening}
\begin{equation}\label{Channel Hardening_CO_INAC}
\mathop {\lim }\limits_{L \to \infty } {\left| {\tilde h} \right|^2} = {\left( {\sum\limits_{l = 1}^L {\left| {{\beta _l}{h_{{\rm{r}},l}}{g_{{\rm{r}},l}}} \right|} } \right)^2} \to {L^2}\beta _l^2m_1^2m_2^2 = m_3^2.
\end{equation}

Therefore, the SINR of both multi-cast and uni-cast in the CO-INAC scenario can be evaluated as follows:
\begin{equation}\label{SNR CO_INAC_navigation}
\mathop {\lim }\limits_{L \to \infty } SIN{R_{{\rm{CO,N}}}} = \frac{{\alpha _M^2}}{{\alpha _U^2}},
\end{equation}
and
\begin{equation}\label{SNR CO_INAC_communication}
\mathop {\lim }\limits_{L \to \infty } SIN{R_{{\rm{CO,C}}}} = \frac{{\alpha _U^2m_3^2l\left( d \right)l\left( {{d_{{\rm{RU}}}}} \right){{g_{{\rm{sp}}}}}p}}{{{\rho ^2}}},
\end{equation}
respectively.

To do so, the channel capacity of multi-cast and uni-cast in the CO-INAC scenario can be evaluated based on the following propositions.
\begin{proposition}\label{Expected rate CO_INAC}
\emph{Based on~\eqref{SNR CO_INAC_navigation} and~\eqref{SNR CO_INAC_communication}, the channel capacity of the multi-cast and uni-cast in the CO-INAC scenario can be evaluated, which are given by}
\begin{equation}\label{rate CO_INAC_navigation}
\mathop {\lim }\limits_{L \to \infty } {R_{{\rm{CO,N}}}} = {\log _2}\left( {1 + \frac{{\alpha _M^2}}{{\alpha _U^2}}} \right),
\end{equation}
and
\begin{equation}\label{rate CO_INAC_communication}
\mathop {\lim }\limits_{L \to \infty } {R_{{\rm{CO,C}}}} = {\log _2}\left( {1 + \frac{{\alpha _U^2m_3^2l\left( d \right)l\left( {{d_{{\rm{RU}}}}} \right)p{{g_{{\rm{sp}}}}}}}{{{\rho ^2}}}} \right),
\end{equation}
respectively.
\end{proposition}

\begin{remark}\label{remark5:channel capacity of CO-INAC}
In the CO-INAC scenario, the results of~\eqref{rate CO_INAC_navigation} demonstrate that the communication performance is better than that of the navigation performance in the high-SNR regime or the case that the number of RIS elements is high enough.
\end{remark}

\begin{remark}\label{remark6:channel capacity}
The results of~\eqref{rate CO_INAC_communication} also indicate that the channel capacity of navigation is a constant in the CO-INAC scenario. It is also demonstrated that the CO-INAC scenario should be applied to the ``Communication-first-scenario'', while navigation is an integrated function.
\end{remark}

\subsection{Navigation Performance}

Positional accuracy has been required from 10 cm in 2-dimension (2D) space in the fifth generation era to 1 cm in 3-dimension space in the 6G~\cite{t2}, which cannot be satisfied by the conventional GNSS network. In the proposed NOMA-RIS-aided INAC networks, since the channel capacity of navigation is much higher than the conventional GNSS networks, the positional accuracy can be boosted. In order to evaluate the position of the INAC-user, we first define the actual distance ${r_{\tau i}}$ between satellite ${s_i}$ and the INAC-user as
\begin{equation}\label{e119}
{r_{\tau i}} = {\rho _{ci}} - c\Delta{t_r},
\end{equation}
where $\Delta{t_r}$ denotes the clock error between the INAC-user and standard clock. ${\rho _{ci}}$ represents the pseudo-range between the INAC-user and satellite.
Therefore, the navigation matrix for the INAC-user can be given by
\begin{equation}\label{e20}
\begin{array}{*{20}{c}}
{\sqrt {{{\left( {{x_1} - {x_u}} \right)}^2} + {{\left( {{y_1} - {y_u}} \right)}^2} + {{\left( {{z_1} - {z_u}} \right)}^2}}  = {\rho _{c1}} - c{\rm{\Delta }}{t_r}}\\
{\sqrt {{{\left( {{x_2} - {x_u}} \right)}^2} + {{\left( {{y_2} - {y_u}} \right)}^2} + {{\left( {{z_2} - {z_u}} \right)}^2}}  = {\rho _{c2}} - c{\rm{\Delta }}{t_r}}\\
{\sqrt {{{\left( {{x_3} - {x_u}} \right)}^2} + {{\left( {{y_3} - {y_u}} \right)}^2} + {{\left( {{z_3} - {z_u}} \right)}^2}}  = {\rho _{c3}} - c{\rm{\Delta }}{t_r}}\\
{r_{\tau R}} + \sqrt {{{({x_R} - {x_u})}^2} + {{({y_R} - {y_u})}^2}{\rm{ + }}{{({z_R} - {z_u})}^2}} {\rm{ = }}{\rho _{c4}} - c\Delta{t_r},
\end{array}
\end{equation}
where $x_u$, $y_u$ and $z_u$ denote the location of x-axis, y-axis and z-axis of the INAC-user. ${r_{\tau R}}$ denotes the distance between the INAC-user and RIS, which can be given by:
\begin{equation}\label{e21}
{r_{\tau R}} = \sqrt {{{({x_4} - {x_R})}^2} + {{({y_4} - {y_R})}^2}{\rm{ + }}{{({z_4} - {z_R})}^2}},
\end{equation}
where $x_R$, $y_R$ and $z_R$ denote the location of the RIS.

We then solve the positioning equations using the Least Squares Method (LSM) algorithm, which is a key performance indicator for the INAC networks. Hence, we define that ${\bf {B}}$ is the observation vector. Note that each element of ${\bf {B}}$ can be given by:
\begin{equation}\label{e34444}
{ {b}}_i = {\rho _{ci}}\left( {{{\bf{x}}_r}} \right) - {\rho _{ci}}\left( {{{\bf{x}}_{r0}}} \right).
\end{equation}\par

${\bf{V}}$ denotes the relation vector, which is divided into ${{\bf{u}}_i}$ and $d{{\bf{x}}_0}$, and they can be expressed respectively as

\begin{equation}
{{\bf{u}}_i} = \left[ {\begin{array}{*{20}{c}}
{\frac{{{x_i} - {x_{u0}}}}{{{r_i}\left( {{{\bf{x}}_{r0}}} \right)}}}&{\frac{{{y_i} - {y_{u0}}}}{{{r_i}\left( {{{\bf{x}}_{r0}}} \right)}}}&{\frac{{{z_i} - {z_{u0}}}}{{{r_i}\left( {{{\bf{x}}_{r0}}} \right)}}}&1
\end{array}} \right],
\end{equation}
\begin{equation}
d{{\bf{x}}_0} = \left[ {\begin{array}{*{20}{c}}
{{x_u} - {x_{u0}}}\\
{{y_u} - {y_{u0}}}\\
{{z_u} - {z_{u0}}}\\
{{\rm{c}}\left( {{\rm{\Delta }}{t_r} - {\rm{\Delta }}{t_{r0}}} \right)}
\end{array}} \right].
\end{equation}

 ${\bf{X}}$ is the location of the INAC-user and its clock error, whose elements denote the difference between the initial solution and the current solution.
Hence, from (\ref{e34444}), the positioning equation for satellite ${s_i}$ can be expressed as
\begin{equation}\label{equa2}
b_i = {{\bf{u}}_i}d{{\bf{x}}_0} + {N_0}.
\end{equation}

Note that $d{{\bf{x}}_0}$ is the location of the satellite and its clock error. We then set ${\bf{U}} = {\left[ {\begin{array}{*{20}{c}}
{{\bf{u}}_1^T}&{{\bf{u}}_2^T}&{{\bf{u}}_3^T}&{{\bf{u}}_4^T}
\end{array}} \right]^T}$ is systematic transformation relationship vector. \par

When the system noise ${N_0}$ is neglected, solution $d{{\bf{x}}_{r0}}$ of the equation is given by
\begin{equation}\label{equa3}
d{{\bf{x}}_{r0}} = {({{\bf{U}}^T}{\bf{U}})^{ - 1}}{{\bf{U}}^T}{\bf B},
\end{equation}
where $d{{\bf{x}}_{r0}}$ is the difference between current solution ${{\bf{x}}_{r1}}$ and initial solution ${{\bf{x}}_{r0}}$, and the current solution ${{\bf{x}}_{r1}}$ can be expressed as
\begin{equation}
{{\bf{x}}_{r1}} = {{\bf{x}}_{r0}} + d{{\bf{x}}_{r0}},
\end{equation}
where ${{\bf{x}}_{r1}}$ can be used as the initial solution for the next iteration.

Then, with the number of iterations ${iters}$ and minimum cost function value $loss$, and the calculation process is expressed in~{\bf{Algorithm 1}}.

\begin{algorithm}[H]
\caption{LSM Position Solution} \label{Algorithm 1}
\begin{algorithmic}[1]
\REQUIRE

$\bf{B}$,$\bf{V}$,${{\bf{X}}_0}$\\
%\ENSURE
%$\mathbf{X}$\\
\STATE Initialize ${{\bf{X}}_0}$\\
    \IF { first position }
            \STATE ${{\bf{X}}_0} = {\left[ {\begin{array}{*{20}{c}}
0&0&0&0
\end{array}} \right]^T}$
    \ELSE
        \STATE ${{\bf{X}}_0} = {\left[ {\begin{array}{*{20}{c}}
{{x_{u0}}}&{{z_{u0}}}&{{y_{u0}}}&0
\end{array}} \right]^T}$
    \ENDIF
    \FOR { $i=1$ to $ {iters}$ }
        \STATE compute ${\bf{P}}({\bf{x}})$ by formula:\ \STATE ${\bf{P}}({\bf{x}}) = {({\bf{B}} - {\bf{VX}})^T}({\bf{B}} - {\bf{VX}})$\\
            \IF {${\bf{P}}({\bf{x}}) < loss$}
                \STATE break
            \ENDIF
            \STATE compute $\mathbf{X}$ by formula: ${\bf{X}} = {{\bf{X}}_0} + {{\bf{V}}^{ - 1}}{\bf{B}}$\\
            \STATE compute ${{\bf{X}}_0}$ by formula: ${{\bf{X}}_0} = {\bf{X}}$\\
    \ENDFOR
%\STATE \textbf{return} $\mathbf{X}$
\ENSURE

$\mathbf{X}$\\
\end{algorithmic}
\end{algorithm}

Then, by utilizing {\bf Algorithm~\ref{Algorithm 1}}, the position of the INAC-user can be solved.

\section{Performance Analysis of the NO-INAC scenario}
Since the channel statistics and navigation performance of both CO-INAC and NO-INAC scenarios are identical, we just analyse SINRs, OPs, and channel capacity of the NO-INAC scenarios.

\subsection{SINR of the NO-INAC scenario}

Similarly, in the NO-INAC scenarios, since more power is allocated to uni-cast signals, which should be detected in the beginning with the SINR
\begin{equation}\label{SINR_NO_communication}
SIN{R_{{\rm{NO,C}}}} = \frac{{\alpha _U^2{{\left| {\tilde h} \right|}^2}l\left( d \right)l\left( {{d_{{\rm{RU}}}}} \right)p {{g_{{\rm{sp}}}}}  }}{{\alpha _M^2{{\left| {\tilde h} \right|}^2}l\left( d \right)l\left( {{d_{{\rm{RU}}}}} \right)p{{g_{{\rm{sp}}}}} + {\rho ^2}}}.
\end{equation}

Therefore, the SINR for detecting the multi-cast signal is given by:
\begin{equation}\label{SINR_NO_navigation}
SIN{R_{{\rm{NO,N}}}} = \frac{{\alpha _M^2{{\left| {\tilde h} \right|}^2}l\left( d \right)l\left( {{d_{{\rm{RU}}}}} \right)p {{g_{{\rm{sp}}}}} }}{{{\rho ^2}}}.
\end{equation}

\subsection{Outage Probability}
The INAC-user first detects the signal of the uni-cast with the OP defined as follows:
\begin{equation}\label{OP_NO_INAC_detect CO_signal}
{P_{{\rm{NO,C}}}} = \mathbb{P}  \left( {{{\log }_2}(1 + SIN{R_{{\rm{NO,C}}}}) < {R_U}} \right).
\end{equation}

Then, the INAC-user will decode the multi-cast signal, and the OP is defined as follows:
\begin{equation}\label{OP_NO_INAC_detect NO_signal}
\begin{array}{l}
{P_{{\rm{NO,N}}}} =\mathbb{P} \left( {{{\log }_2}(1 + SIN{R_{{\rm{NO,C}}}}) < {R_U}} \right)\\
 + \mathbb{P}\left( {{{\log }_2}(1 + SIN{R_{{\rm{NO,C}}}}) > {R_U}, }\right. \\
\left.{  {{\log }_2}(1 + SIN{R_{{\rm{NO,N}}}}) < {R_M}} \right),
\end{array}
\end{equation}

\begin{theorem}\label{Theorem3:Outage NO_INAC communication}
\emph{In the NO-INAC scenarios, assuming that $ {\alpha _U^2 - \alpha _M^2{\varepsilon _U}}>0$, the closed-form OP expression of the INAC-user for decoding the uni-cast signal can be expressed as}
\begin{equation}\label{outage analytical results NO-INAC decode unicast in theorem3}
\begin{aligned}
{P_{{\rm{NO,C}}}} =  \frac{1}{2}\left( {{\rm{erf}}\left( {\frac{{m_3}+ {\sqrt {{\omega _{{\rm{NO,C}}}}} }}{{\sqrt {2{v_3}} }}} \right) }\right.\\
\left.{ - {\rm{erf}}\left( {\frac{{m_3}+{\sqrt {{\omega _{{\rm{NO,C}}}}}  }}{{\sqrt {2{v_3}} }}} \right)} \right),
\end{aligned}
\end{equation}
\emph{where ${\omega _{{\rm{NO,C}}}} = \frac{{{\varepsilon _U}{\rho ^2}}}{{\left( {\alpha _U^2 - \alpha _M^2{\varepsilon _U}} \right)\gamma }}$.}
\begin{proof}
It is easy to prove Theorem~\ref{Theorem3:Outage NO_INAC communication} in the same way as Appendix B.
\end{proof}
\end{theorem}

\begin{theorem}\label{Theorem4:Outage NO-INAC decode navigation signal closed form by error func}
\emph{In the NO-INAC scenarios, assuming that $ {\alpha _U^2 - \alpha _M^2{\varepsilon _U}}>0$, the closed-form OP expression for decoding the multi-cast signal and uni-cast signal can be expressed as}
\begin{equation}\label{outage analytical results NO-INAC decode multicast in theorem3}
\begin{aligned}
{P_{{\rm{NO,N}}}} =  \frac{1}{2}\left( {{\rm{erf}}\left( {\frac{{\sqrt {{m_3}+{\omega _{{\rm{NO,N}}}}}  }}{{\sqrt {2{v_3}} }}} \right) }\right.\\
\left.{ - {\rm{erf}}\left( {\frac{{m_3}+{\sqrt {{\omega _{{\rm{NO,N}}}}}  }}{{\sqrt {2{v_3}} }}} \right)} \right),
\end{aligned}
\end{equation}
\emph{where ${\omega _{{\rm{NO,N}}}} = \max \left\{ {\frac{{{\varepsilon _M}{\rho ^2}}}{{\alpha _M^2\gamma }},\frac{{{\varepsilon _U}{\rho ^2}}}{{\left( {\alpha _U^2 - \alpha _M^2{\varepsilon _U}} \right)\gamma }}} \right\}$.}
\begin{proof}
It is easy to prove Theorem~\ref{Theorem4:Outage NO-INAC decode navigation signal closed form by error func} in the same way as Appendix B.
\end{proof}
\end{theorem}

\begin{corollary}\label{corollary3:Outage NO_INAC_decode communication user asymptotic}
\emph{Assuming that $ {\alpha _U^2 - \alpha _M^2{\varepsilon _U}}>0 $, the OP of the INAC-user for decoding the uni-cast signal in NO-INAC scenarios can be approximated in closed form by}
\begin{equation}\label{Corro3:NO_INAC_decoding communication}
\begin{aligned}
{{\bar P}_{{\rm{NO,C}}}} = &  \frac{4}{{\sqrt \pi  }}\sum\limits_{n = 0}^\infty  {\frac{{{{\left( { - 1} \right)}^n}}}{{n!\left( {2n + 1} \right){{\left( {2{v_3}} \right)}^{\frac{{2n + 1}}{2}}}}}} \\
& \times \sum\limits_{k = 1,3...2n + 1}^{} {\left( {\begin{array}{*{20}{c}}
{2n + 1}\\
k
\end{array}} \right)} m_3^{2n + 1 - k}\omega _{{\rm{NO,C}}}^{\frac{k}{2}},
\end{aligned}
\end{equation}
\begin{proof}
Similar to Appendix C, the proof is complete.
\end{proof}
\end{corollary}

\begin{corollary}\label{corollary4:Outage NO_INAC_decode navigation user asymptotic}
\emph{Assuming that $ {\alpha _U^2 - \alpha _M^2{\varepsilon _U}}>0$, the OP of the INAC-user for decoding multi-cast and uni-cast signals in NO-INAC scenarios can be approximated in closed form by}
\begin{equation}\label{Corro4:NO_INAC_decoding navigation}
\begin{aligned}
{{\bar P} _{{\rm{NO,N}}}} = & \frac{4}{{\sqrt \pi  }}\sum\limits_{n = 0}^\infty  {\frac{{{{\left( { - 1} \right)}^n}}}{{n!\left( {2n + 1} \right){{\left( {2{v_3}} \right)}^{\frac{{2n + 1}}{2}}}}}} \\
& \times \sum\limits_{k = 1,3...2n + 1}^{} {\left( {\begin{array}{*{20}{c}}
{2n + 1}\\
k
\end{array}} \right)} m_3^{2n + 1 - k}\omega _{{\rm{NO,N}}}^{\frac{k}{2}}.
\end{aligned}
\end{equation}
\begin{proof}
Similar to Corollary~3, the proof can be readily completed.
\end{proof}
\end{corollary}

\begin{proposition}\label{proposition1: v diversity order}
\emph{Based on \textbf{Corollary~\ref{corollary3:Outage NO_INAC_decode communication user asymptotic}} and \textbf{Corollary~\ref{corollary4:Outage NO_INAC_decode navigation user asymptotic}}, the diversity order of the INAC-user in the NO-INAC scenarios are given by}
\begin{equation}\label{diversity order of w}
{d_{{\rm{NO,C}}}}= {d_{{\rm{NO,N}}}} =  - \mathop {\lim }\limits_{\frac{p}{{{\sigma ^2}}} \to \infty } \frac{{\log {{\bar P}_{{\rm{NO,C}}}}}}{{\log \frac{p}{{{\sigma ^2}}}}} \approx m_3.
\end{equation}
\end{proposition}

\subsection{Channel Capacity}

Similarly, the channel capacity of multi-cast and uni-cast in the NO-INAC scenarios can be evaluated based on the following propositions. Hence, when the number of RIS elements is high enough, the SINR of uni-cast and multi-cast can be approximated to
\begin{equation}\label{High SNR NO_INAC_Communication}
\mathop {\lim }\limits_{L \to \infty } SIN{R_{{\rm{NO,C}}}} = \frac{{\alpha _U^2}}{{\alpha _M^2}},
\end{equation}
and
\begin{equation}\label{High SNR NO_INAC_Navigation}
\mathop {\lim }\limits_{L \to \infty } SIN{R_{{\rm{NO,N}}}} = \frac{{\alpha _M^2m_3^2l\left( d \right)l\left( {{d_{{\rm{RU}}}}} \right)p {{g_{{\rm{sp}}}}}  }}{{{\rho ^2}}}
\end{equation}

\begin{proposition}\label{Expected rate NO_INAC}
\emph{Based on~\eqref{High SNR NO_INAC_Communication} and~\eqref{High SNR NO_INAC_Navigation}, the channel capacity of multi-cast and uni-cast in the NO-INAC scenarios can be evaluated, which are given by}
\begin{equation}\label{Rate_NO_INAC_Communications}
\mathop {\lim }\limits_{L \to \infty } {R_{{\rm{NO,C}}}} = {\log _2}\left( {1 + \frac{{\alpha _U^2}}{{\alpha _M^2}}} \right),
\end{equation}
and
\begin{equation}\label{Rate_NO_INAC_NAvigation}
\mathop {\lim }\limits_{L \to \infty } {R_{{\rm{NO,N}}}} = {\log _2}\left( {1 + \frac{{\alpha _M^2m_3^2l\left( d \right)l\left( {{d_{{\rm{RU}}}}} \right)p{{g_{{\rm{sp}}}}}}}{{{\rho ^2}}}} \right),
\end{equation}
respectively.
\end{proposition}

\begin{remark}\label{remark7:NO-INAC}
Similar to \textbf{Remark~\ref{remark6:channel capacity}}, the results in this Section indicate that the channel capacity of communication is a constant in the NO-INAC scenario, which demonstrated that the NO-INAC scenario should be applied to the ``navigation-first-scenario''.
\end{remark}

\section{Numerical Studies}

Here, we present numerical results regarding the performance evaluation of NOMA-RIS-aided INAC networks. The accuracy of our analytical results is verified through Monte Carlo simulations. Note that since the INAC-user has more functions than that of navigation user, we only focus on the performance of the INAC-user. We have set the bandwidth to 30 MHz and the carrier frequency to 10 GHz. The power of the AWGN is set to ${{\rho ^2}}= 174+ 10 {\rm log}_{10}(BW)$ dBm. In this case, the Rician fading parameter indicates that the LoS and NLoS links are represented by ${\cal K} = 0$ and ${\cal K} > 0$, respectively. The target rates are $R_M=0.0005$ and $R_U=0.001$ bits per Hz per second (bps/Hz).
The power allocation factors are set to $\alpha_M^2=0.6$ and $\alpha_U^2=0.4$ in the CO-INAC scenario. On the contrary, in the NO-INAC scenario, the power allocation factors are set to $\alpha_U^2=0.9$ and $\alpha_M^2=0.1$.
The radius of the earth and the radius of MEO satellites are set to 6378 km and 20000 km, respectively. The elevation angle of the INAC satellite is set to $ \pi/10$ unless otherwise stated. The distance of the RIS-user links is set to $d_{\rm RU}=10$m. Except where otherwise indicated, the path loss exponents for satellite-RIS and RIS-user connections are ser to $\alpha_1=2$ and $\alpha_2=2.2$, respectively.

\begin{figure}[t!]
\centering
\includegraphics[width =3in]{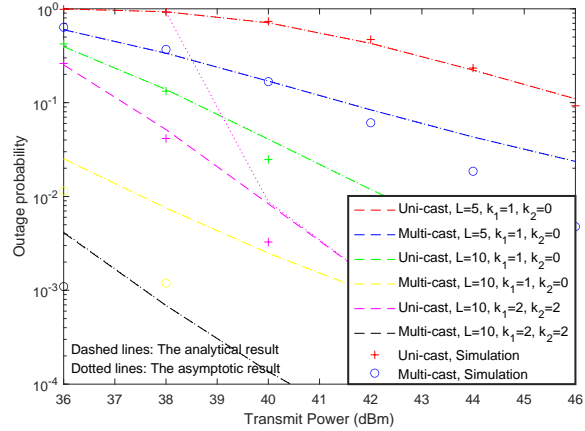}
\caption{OP of the NOMA-RIS-aided CO-INAC scenario versus the transmit power.}
\label{CO_INAC_OP_power_fig1}
\end{figure}

\emph{1) Impact of the Transmit Power on OP:} Fig.~\ref{CO_INAC_OP_power_fig1} depicts the OP of the NOMA-RIS-aided INAC networks. Analytical and asymptotic results are represented by dashed curves and dotted curves, respectively. We can observe that when the number of RIS elements serving the INAC-user increases, the OP decreases. Due to increased diversity order, receiving signal power can be significantly increased when more RISs are employed. The results also indicate that the slope of the curves increases with the increased number of RIS elements and with the increased fading parameters, which validates our~\textbf{Remark~\ref{remark2:impact of fading environment on diversity order}}. Moreover, since the asymptotic results are derived based on the Taylor expansion, the asymptotic results may not be applicable when the number of RIS elements is large enough, which verifies our~\textbf{Remark\ref{remark2:Asymptotic condition}}. Based on the comparison between the analytical results and simulations, we can also observe that the proposed analytical results based on the CLT may perform better in the low-SNR regimes when the number of RIS elements is not high enough.

\begin{figure}[t!]
\centering
\includegraphics[width = 3 in]{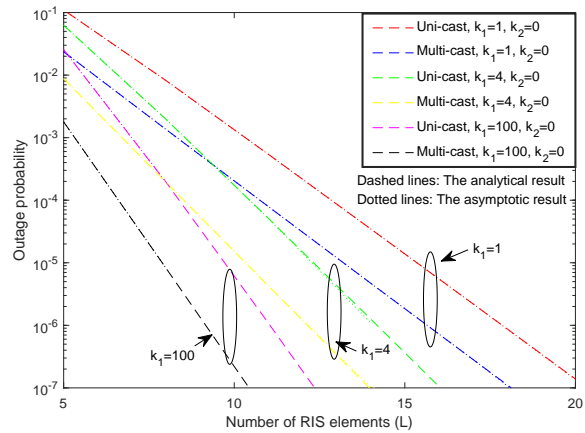}
\caption{OP of the NOMA-RIS-aided CO-INAC scenario versus the number of RIS elements parameterized by the fading factors. The transmit power is set to 46 dBm.}
\label{CO-INAC-impact on L}
\end{figure}

\emph{2) Impact of the Number of RIS Elements on OP:} In Fig.~\ref{CO-INAC-impact on L}, we evaluate the OP of the INAC-user in the CO-INAC scenarios with the different number of RIS elements. On the one hand, as expected, as the number of RIS elements increases, the OP decreases. On the other hand, the fading parameters also impact the OP of both multi-cast and uni-cast signals for the INAC-user, which verify our~\textbf{Remark~\ref{remark2:impact of fading environment on diversity order}}.
observe that when the fading parameter of the satellite-RIS approaches infinity, the diversity order of the INAC-user is mainly impacted by the RIS-user link, which verifies~\textbf{Remark~\ref{remark3:strong los impact}}. Note that this phenomenon is applicable in the MEO-INAC networks, which is due to the fact that the right-handed circularly polarized wave is utilized, and the NLoS links between satellite and RIS are usually mitigated.
We can also see that the asymptotic results of the black and purple curves do not exist in the cast of ${\mathcal{K}_r}=100$, which is due to the fact that the Taylor expansion does not exist when the number of RIS elements or fading parameters are too high. This also verifies our~\textbf{Remark~\ref{remark2:Asymptotic condition}}.

\begin{figure}[t!]
\centering
\includegraphics[width =3 in]{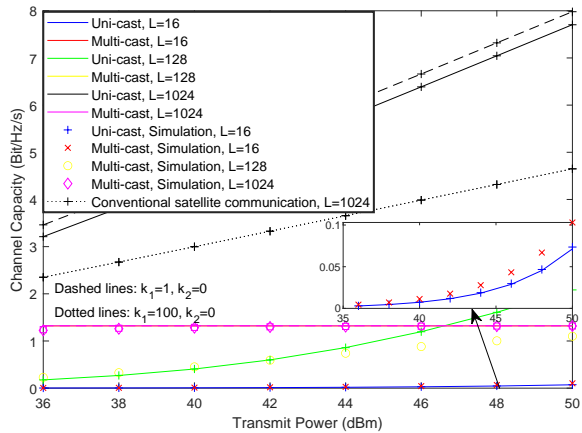}
\caption{Channel capacity of the NOMA-RIS-aided CO-INAC scenario versus the transmit power parameterized by the number of RIS elements.}
\label{Channel Capacity_impact power_fig3}
\end{figure}

\emph{3) Impact of the Transmit Power on Channel Capacity:} Let us now study the performance of channel capacity in Fig.~\ref{Channel Capacity_impact power_fig3}. Based on this analysis, it is evident that there is a substantial gap between approximated results and simulation results. This is due to the fact that the simulation results are derived when the number of RIS elements is high enough. However, when the number of RIS elements is high enough, we have a close agreement between the simulation and approximated results. Observe that the channel capacity of the uni-cast signal increases as the transmit power increases, whereas the channel capacity of the multi-cast approaches a constant, which indicates that the CO-INAC scenario can provide better performance for communication when the number of RIS elements is high enough. This verified the insights gleaned from \textbf{Remark~\ref{remark5:channel capacity of CO-INAC}}. Based on the comparison between blue, green and black curves, more RIS elements are expected to improve the channel capacity of the proposed INAC networks. It is also indicated that when the number of RIS elements is high enough, the channel capacity of satellite communications can be even higher than that of civil communications. In addition, based on the black curves and dashed curves, the fading parameters also improve the channel capacity for the uni-cast signal, which indicates that the LoS links are expected between the INAC satellite and RIS. We simply adopt the counterpart of NOMA, time-division multiple access (TDMA), to better compare the performance of the conventional satellite communications and INAC networks. As we can observe from the dotted and dashed black curves, it is indicated that the performance of the INAC user is better than that of the conventional satellite communications in the high-SNR regimes, which illustrates the performance gain of the INAC networks.

\begin{figure}[t!]
\centering
\includegraphics[width =3 in]{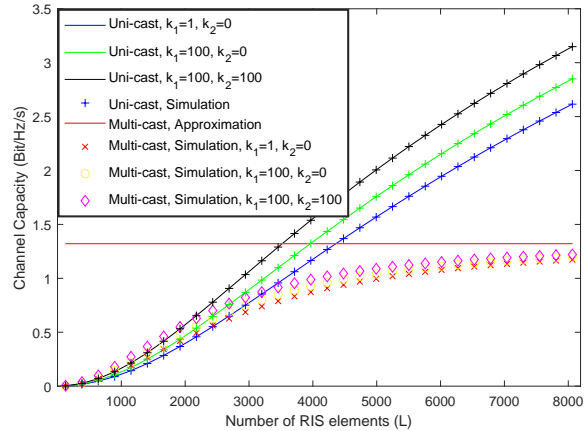}
\caption{Channel capacity of the NOMA-RIS-aided CO-INAC scenario versus the number of RIS elements. The spread spectrum gain is set to 30 dB.}
\label{Channel Capacity_impact L_fig4}
\end{figure}

\emph{4) Impact of the Number of RIS Elements on Channel Capacity:} Fig.~\ref{Channel Capacity_impact L_fig4} compares the channel capacity of the multi-cast and uni-cast signals versus the number of RIS elements parameterized by the fading coefficients. In order to illustrate the performance gain of RIS, the spread spectrum gain is set to 30 dB. As can be observed, the LoS of both the satellite-RIS and the RIS-user links increase the channel capacity of the INAC user. As a result of simulation results and approximated results of multicast, we have verified the accuracy of our conclusions. In addition, when more RIS elements are utilized, the channel capacity of the INAC-user can be significantly increased. This is due to the fact that the spatial diversity gain can be significantly increased by increasing the number of RIS elements.
%
%\begin{table}
%\caption{\\ DIVERSITY ORDER AND HIGH-SNR SLOPE}
%\centering
%\begin{tabular}{|l|c|c|c|}
%\hline
%Access Mode & Rx & D & S \\
%\hline
%\multirow{2}{*}{RIS-aided NOMA}
%& $W$ & $\frac{N {m_1} {m_W}W}{m_1 +{m_W}} $ & 1 \\
%\cline{2-4}
%& $v$ & $\frac{N {m_1} {m_v}v}{m_1 +{m_v}} $ & 0 \\
%\hline
%\multirow{2}{*}{Conventional NOMA}
%& $W$ & $W$ & 1 \\
%\cline{2-4}
%& $v$ & $v$ & 0 \\
%\hline
%\multirow{2}{*}{OMA}
%& $W$ & $W$ & 0.5 \\
%\cline{2-4}
%& $v$ & $v$ & 0.5 \\
%\hline
%\end{tabular}
%\label{DIVERSITY ORDER AND HIGH SNR SLOPE FOR RIS Networks}
%\end{table}
%
%
\begin{figure}[t!]
\centering
\includegraphics[width =3 in]{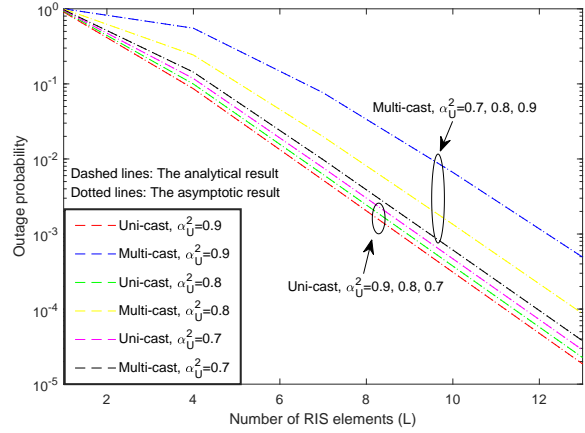}
\caption{OP of the NOMA-RIS-aided NO-INAC scenario versus the number of RIS elements and the power allocation factors. The fading parameters are set to ${\mathcal{K}}_r=1$ and ${\mathcal{K}}_g=0$.}
\label{NO_INAC_Outage_fig5_compare with CO_INAC}
\end{figure}

\emph{5) Impact of the Power Allocation Factors on OP:} In Fig.~\ref{NO_INAC_Outage_fig5_compare with CO_INAC}, we then evaluate the outage of our NOMA-RIS-aided NO-INAC scenarios. We can see that allocating more power to the uni-cast can improve the outage performance of communication, whereas the outage performance of navigation decreases. More specifically, the outage performance of the navigation improves more dramatically than that of the communication, which indicates that it is preferable to allocate more power to the multi-cast for enhancing the outage performance in the NO-INAC scenarios.

\begin{figure}[t!]
\centering
\includegraphics[width =3 in]{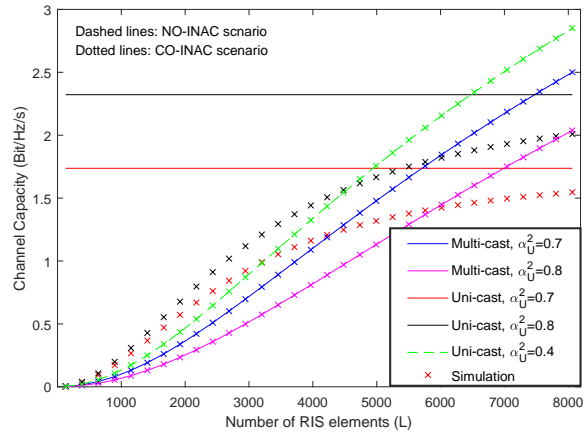}
\caption{Channel capacity of the proposed NOMA-RIS-aided NO-INAC scenario versus the number of RIS elements. The fading parameters are set to ${\mathcal{K}}_r=100$ and ${\mathcal{K}}_g=0$. The results of CO-INAC are calculated from Fig.~\ref{Channel Capacity_impact L_fig4}. }
\label{Rate_compare_with_CO_INAC_FIG6}
\end{figure}

\emph{6) Comparing the CO-INAC Scenario to the NO-INAC Scenario:} In Fig.~\ref{Rate_compare_with_CO_INAC_FIG6}, we evaluate the channel capacity of the proposed NOMA-RIS-aided NO-INAC scenario as well as that of its CO-INAC counterparts, which is provided as the benchmark from Fig.~\ref{Channel Capacity_impact L_fig4}. On the one hand, observe from the figure that more power allocated to the uni-cast in the NO-INAC scenarios can improve the upper bound of the uni-cast signal. On the other hand, the channel capacity of the multi-cast signal in the NO-INAC scenarios is much higher than that of the CO-INAC scenarios, which indicates that the NO-INAC scenario may provide higher accuracy for navigation. Note that in conventional GNSS, the pseudo-range is obtained by the spread spectrum code due to the severe path loss. However, with the aid of RIS in the NO-INAC scenarios, the navigation accuracy can be improved by employing other techniques, e.g., the ultra-wideband (UWB) navigation technique. Since the outage requirement of communication is usually higher than that of the navigation, we also compare the performance of the uni-cast in both CO-INAC and NO-INAC scenarios. Based on the comparison between the CO-INAC and NO-INAC scenarios, we can see that a cross point of curves occurs, which indicates that NOMA-RIS-aided hybrid CO/NO-INAC scenario may be a good solution, where two scenarios can be alternatively changed based on the number of RIS elements, which provides more engineering insights for practice.

\emph{7) Minimal Number of Satellites:}
Based on the elevation angle $\vartheta $, the geocentric angle of ground user can be given by:
\begin{equation}\label{geocentric angle}
\upsilon  = \arccos \left( {\frac{{{r_e}}}{{{r_e} + {r_m}}}\cos \vartheta } \right) - \vartheta .
\end{equation}
Then, the coverage area can be evaluated as:
\begin{equation}\label{coverage area}
{\rm{A}} = 2\pi r_e^2\left( {1 - \cos \upsilon } \right).
\end{equation}

\begin{figure}[t!]
\centering
\includegraphics[width =3in]{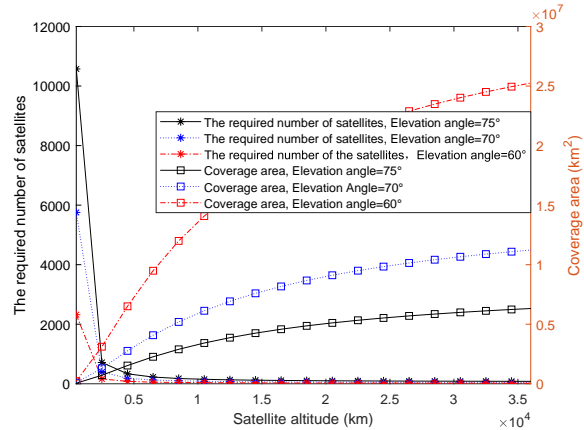}
\caption{The minimal required number of satellites versus the radius of the satellites as well as the elevation angle.}
\label{required number of satellite fig7}
\end{figure}

By doing so, In Fig.~\ref{required number of satellite fig7}, we evaluate the minimal required number of satellites. The results indicate that when the radius of satellites is lower than 500 km, the minimal required number of satellites is greater than 10000, which is not cost-effective. Moreover, based on the insights from Starlink, the life cycle of LEO satellites is lower than expected, resulting in the waste of equipment and orbit. Therefore, if the radius of satellites is increased to 20000 km, only 40 satellites are needed for global wireless services. Note that similar to the civil communications, due to the fact that the overlaps of multiple satellites occur in practice, hence requiring more satellites. However, this is beyond the scope of this treatise. We can also observe that with the increased elevation angle, the coverage area is decreased. This is due to the fact that the coverage area is affected by the angle of the main lobe.

%The results of the FD-relay and HD-relay are given by ${\bar R}_{{\rm{F}},v}+{\bar R}_{{\rm{F}},W} $ and ${\bar R}_{{\rm{H}},v}+ {\bar R}_{{\rm{H}},W} $, respectively. The transmit power of the HD-relay and FD-relay are set to $p_d=(p-10)$ dBm. We can see that the network throughput gap between the RIS-aided NOMA network and the other pair of relay aided networks becomes smaller, when the number of RISs is increased. Observe that for the case of $N=18$, the proposed RIS-aided NOMA network is capable of outperforming both the FD and HD relay aided networks, which indicates that the RIS-aided NOMA network becomes more competitive, when the number of RISs is high enough.

\begin{figure}[t!]
\centering
\includegraphics[width =3 in]{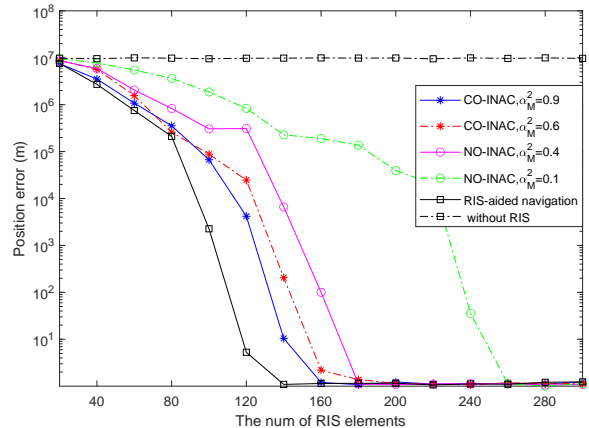}
\caption{Navigation accuracy of both CO-INAC and NO-INAC scenarios versus the number of RIS elements with different power allocation factors.}
\label{navigation accuracy fig8}
\end{figure}

\emph{8) Navigation Accuracy of the CO-INAC and NO-INAC Scenarios:} Fig.~\ref{navigation accuracy fig8} evaluates the navigation accuracy of the proposed NOMA-RIS-aided INAC networks versus the number of RIS elements. Both the CO-INAC and NO-INAC scenarios are evaluated to better illustrate the performance of navigation. On the one hand, we can see that the navigation performance of the NO-INAC scenario is better than that of the CO-INAC scenario when the number of RIS elements is small. This is due to the fact that more power is allocated to navigation.
On the other hand, it can be observed that the navigation accuracy of both CO-INAC and NO-INAC scenarios are identical when the number of RIS elements is high enough. This is due to the fact that conventional navigation is based on the spread coding technique, which cannot provide additional navigation accuracy in the high-SNR regimes. As we can see from the black curves, the performance of the RIS-aided navigation is better than that of the INAC networks when the number of RIS elements is not high enough. Furthermore, we can also observe that when the number of RIS elements is high enough, the performance of RIS-aided navigation, CO-INAC and NO-INAC are nearly identical. In addition, we can also observe from the curve without RIS, the position error approaches infinity due to the fact that the satellite-user link does not exist.

\section{Conclusions}

We first reviewed the recent advances in satellite networks. In order to provide a cost-effective satellite network, the navigation and communication functions were integrated into the MEO satellites. To serve all ground users simultaneously, we adopted the NOMA technique, where the communication and navigation signals were composed. Furthermore, to satisfy the requirement of communications, the RIS technique was adopted for improving the channel capacity. For the purposes of characterization of the performance of the CO-INAC and NO-INAC scenarios, new channel statistics, operational parameters, channel capacity, and navigation accuracy were derived in closed-form.

On the one hand, an important future direction is to extend novel navigation techniques for the proposed INAC networks. Since the channel capacity of the network improved dramatically, the UWB navigation technique may be a good solution. On the other hand, our proposed INAC network can be further extended to the multi-user cases. Based on the insight of~\cite{AI_RIS_R1_4}, artificial intelligence and RIS-aided INAC networks can be integrated for synergistic performance gain, which could be a potential research direction. In addition, since the transmission latency of MEO satellite is too high, it is hard to estimate the perfect CSI~\cite{IpCSI}. Therefore, the imperfect CSI of small-scale channels, the imperfect SIC of NOMA, and the hardware imperfections could also be considered in the future, which are critical for engineering insights.

\numberwithin{equation}{section}
\section*{Appendix~A: Proof of Lemma~\ref{lemma1:new state of effective channel gain}} \label{Appendix:As}
\renewcommand{\theequation}{A.\arabic{equation}}
\setcounter{equation}{0}
In the RIS-aided satellite communication networks, multiple signals may be significantly coherent when the phase and frequency are identical~\cite{Coherence_wave}. Thus, when the reflected SRU signals are co-phased, the INAC-user has the following channel gain:
\begin{equation}\label{first one,appendixA}
{\left| {\tilde h} \right|^2} =  {\left| {{{\bf{G}}_{\rm{r}}}{\bf{\Phi }}{{\bf{H}}_{\rm{r}}}} \right|^2}.
\end{equation}

By properly control the phase shift at RIS, the results of~\eqref{first one,appendixA} can be transformed into~\eqref{effective channel gain before design appen A} as follows:
\begin{equation}\label{effective channel gain before design appen A}
{\left| {\tilde h} \right|^2} = {\left( {\sum\limits_{l = 1}^L {\left| {{\beta _l}{h_{{\rm{r}},l}}{g_{{\rm{r}},l}}} \right|} } \right)^2}.
\end{equation}

We then utilize the central limit theorem to evaluate the channel distribution of the SRU links, where the mean and variance of SR and RU links must be derived. By using the definition of Rician distribution, the channel distribution can be given by ${h_{{\rm{r}},l}} \sim \mathcal{CN} \left( {{m_1},{v_1}} \right)$, which are:
\begin{equation}\label{mean_single channel SR,appendixA}
{m_1} = \sqrt {\frac{\pi }{{4\left( {1 + {{\cal K}_{r,l}}} \right)}}} {}_1{F_1}\left( { - \frac{1}{2},1;{{\cal K}_{r,l}}} \right),
\end{equation}
and
\begin{equation}\label{var_single channel SR,appendixA}
{v_1} = 1 - \frac{\pi }{{4\left( {1 + {{\cal K}_{r,l}}} \right)}}{}_1{F_1}{\left( { - \frac{1}{2},1;{{\cal K}_{r,l}}} \right)^2},
\end{equation}
where ${}_1{F_1}$ is the confluent hypergeometric function of the first kind.

Similarly, the channel distribution ${g_{{\rm{r}},l}}\sim \mathcal{CN}  \left( {{m_2},{v_2}} \right)$ of the RU links is given by
\begin{equation}\label{mean_single channel RU,appendixA}
{m_2} = \sqrt {\frac{\pi }{{4\left( {1 + {{\cal K}_{g,l}}} \right)}}} {}_1{F_1}\left( { - \frac{1}{2},1;{{\cal K}_{g,l}}} \right),
\end{equation}
and
\begin{equation}\label{var_single channel RU,appendixA}
{v_2} = 1 - \frac{\pi }{{4\left( {1 + {{\cal K}_{g,l}}} \right)}}{}_1{F_1}{\left( { - \frac{1}{2},1;{{\cal K}_{g,l}}} \right)^2}.
\end{equation}

By using the central limit theorem and using the random variable property, we can determine the mean and variance of the effective channel gain when the number of RIS elements is sufficient as follows
\begin{equation}\label{appendix first mean}
{m_3} = \mathbb{E} \left\{ {\tilde h} \right\} = {\beta _l}L{m_1}{m_2},
\end{equation}
and
\begin{equation}\label{appendix first variance}
\begin{aligned}
{v_3} &= \mathbb{E} \left\{ {{{\tilde h}^2}} \right\} - \mathbb{E} {\left\{ {\tilde h} \right\}^2}\\
& = {\beta _l}L\left( {m_1^2{v_2} + m_2^2{v_1} + {v_1}{v_2}} \right).
\end{aligned}
\end{equation}

Thus, we can express the PDF of channel gain as:
\begin{equation}\label{appendix channel multiple distribution}
{f_{\left| {\tilde h} \right|}}\left( x \right) = \frac{1}{{\sqrt {2\pi {v_3}} }}\exp \left( { - \frac{{{{\left( {x - {m_3}} \right)}^2}}}{{2{v_3}}}} \right).
\end{equation}

Since the time-domain amplitude must be transformed into the power-domain for evaluating the performance, which can be derived by the Jacobian transform as follows:
\begin{equation}\label{time-domain-amplitude to channel gain}
{f_{\left| {\tilde h} \right|^2}}\left( y \right) = \frac{1}{{2\sqrt y }}\left( {{f_{\left| {\tilde h} \right|}}\left( {\sqrt y } \right) + {f_{\left| {\tilde h} \right|}}\left( { - \sqrt y } \right)} \right).
\end{equation}

Thereby, the power-domain channel gain of the SRU links is given by
\begin{equation}\label{appendix effective channel distribution}
\begin{aligned}
{f_{{{\left| {\tilde h} \right|}^2}}}\left( x \right) = & \frac{1}{{2\sqrt {2\pi {v_3}x} }}\left( {\exp \left( { - \frac{{{{\left( {\sqrt x  + {m_3}} \right)}^2}}}{{2{v_3}}}} \right)  }\right. \\
& + \left.{ \exp \left( { - \frac{{{{\left( {\sqrt x  - {m_3}} \right)}^2}}}{{2{v_3}}}} \right)} \right).
\end{aligned}
\end{equation}

We then derive the CDF of the channel gain, which can be derived by the integration of~\eqref{appendix effective channel distribution} as follows:
\begin{equation}\label{CDF_equation}
\begin{aligned}
{F_{{{\left| {\tilde h} \right|}^2}}}\left( x \right) & = \int_0^y {\frac{1}{{2\sqrt {2\pi {v_3}x} }}\left( {\exp \left( { - \frac{{{{\left( {\sqrt x  + {m_3}} \right)}^2}}}{{2{v_3}}}} \right)  }\right.}\\
& {\left.{ + \exp \left( { - \frac{{{{\left( {\sqrt x  - {m_3}} \right)}^2}}}{{2{v_3}}}} \right)} \right)} dx,\\
&  = \frac{1}{2}\left( {{\rm{erf}}\left( {\frac{{m_3}+{\sqrt x  }}{{\sqrt {2{v_3}} }}} \right) - {\rm{erf}}\left( {\frac{{m_3}-{\sqrt x   }}{{\sqrt {2{v_3}} }}} \right)} \right).
\end{aligned}
\end{equation}

Hence, the PDF and CDF of the channel gain can be derived in Lemma 1, and the proof is complete.

\numberwithin{equation}{section}
\section*{Appendix~B: Proof of Theorem~\ref{Theorem1:Outage CO-INAC decoding communication closed form by error func}} \label{Appendix:Bs}
\renewcommand{\theequation}{B.\arabic{equation}}
\setcounter{equation}{0}

Based on the OP defined in~\eqref{OP_CO_INAC_detect NO_signal}, the OP for decoding the multi-cast signal of the CO-INAC scenario can be rewritten as
\begin{equation}\label{appendix B outage defination}
{P_{{\rm{CO,N}}}} =\mathbb{P} \left( {{{\log }_2}(1 + SIN{R_{{\rm{CO,N}}}}) < {R_M}} \right).
\end{equation}

By some algebraic manipulations,~\eqref{appendix B outage defination} can be formulated as
\begin{equation}\label{appendix B SINR}
{P_{{\rm{CO,N}}}} = \mathbb{P} \left( {{{\left| {\tilde h} \right|}^2} < \frac{{{\varepsilon _M}{\rho ^2}}}{{\left( {\alpha _M^2 - \alpha _U^2{\varepsilon _M}} \right)l\left( d \right)l\left( {{d_{{\rm{RU}}}}} \right)p{g_{{\rm{sp}}}}}}} \right).
\end{equation}

By substituting~\eqref{appendix B SINR} into~\eqref{New Gamma distribution PDF}, the OP is given by
\begin{equation}\label{appendix B outage expression}
{P_{{\rm{CO,N}}}} =  \int_0^{{\omega _{\rm{CO,N}}}} {{f_{{{\left| {\tilde h} \right|}^2}}}\left( x \right)dx}.
\end{equation}
%Thus, the OP of user $W$ can be obtained as
%\begin{equation}\label{Appendix B outage final}
%{P_W} = {\left( {\frac{{\gamma \left( {{k_1},\frac{{{I_{W*}}}}{{{\lambda _1}}}} \right)}}{{\Gamma ({k_1})}}} \right)^{W}}.
%\end{equation}
After some algebraic manipulation, the OP for decoding the multi-cast signal presented by INAC-user in~\eqref{outage analytical results CO-INAC decode multicast in theorem1} can be obtained. The proof is complete.

\numberwithin{equation}{section}
\section*{Appendix~C: Proof of Corollary~\ref{corollary1:Outage CO_INAC_navigation}} \label{Appendix:Cs}
\renewcommand{\theequation}{C.\arabic{equation}}
\setcounter{equation}{0}

In order to glean further engineering insights, we first expand the error function as follows~\cite{Table_of_integrals}:
\begin{equation}\label{Appendix C Lower incomplete gamma expansion}
\begin{aligned}
{\rm{erf}}\left( z \right) = \frac{2}{{\sqrt \pi  }}\sum\limits_{n = 0}^\infty  {\frac{{{{\left( { - 1} \right)}^n}{z^{2n + 1}}}}{{n!\left( {2n + 1} \right)}}} .
\end{aligned}
\end{equation}

Hence the OP of the INAC-user can be expressed as follows:
\begin{equation}\label{Apeendix C first approx}
\begin{aligned}
{{\bar P}_{{\rm{CO,N}}}} =& \frac{1}{{\sqrt \pi  }}\sum\limits_{n = 0}^\infty  {\frac{{{{\left( { - 1} \right)}^n}}}{{n!\left( {2n + 1} \right)}}} \\
& \times \frac{{{{\left( {{m_3} + \sqrt {{\omega _{\rm{C}}}} } \right)}^{2n + 1}} - {{\left( {{m_3} - \sqrt {{\omega _{\rm{C}}}} } \right)}^{2n + 1}}}}{{{{\left( {2{v_3}} \right)}^{\frac{{2n + 1}}{2}}}}}.
\end{aligned}
\end{equation}
Upon involving the binomial expansion and after some algebraic manipulations, the approximate result can be further transformed into
\begin{equation}\label{Appendix C second transform}
\begin{aligned}
{{\bar P}_{{\rm{CO,N}}}} = & \frac{1}{{\sqrt \pi  }}\sum\limits_{n = 0}^\infty  {\frac{{{{\left( { - 1} \right)}^n}}}{{n!\left( {2n + 1} \right){{\left( {2{v_3}} \right)}^{\frac{{2n + 1}}{2}}}}}} \\
& \times \sum\limits_{k = 0}^{2n + 1} {\left( {\begin{array}{*{20}{c}}
{2n + 1}\\
k
\end{array}} \right)} m_3^{2n + 1 - k}\omega _{\rm{C}}^{\frac{k}{2}}\left( {1 - {{\left( { - 1} \right)}^k}} \right).
\end{aligned}
\end{equation}

When $k$ is an odd number, we have $\left( {1 - {{\left( { - 1} \right)}^k}} \right) = 0$. Otherwise, when $k$ is an even number, we have $\left( {1 - {{\left( { - 1} \right)}^k}} \right) = 2$. Then by some algebraic manipulations, the proof is complete. Thus, the results in~\eqref{Corro1:CO_INAC_decoding navigation} can be obtained, and the proof is complete.

\bibliographystyle{IEEEtran}
\bibliography{IEEEabrv,NOMA_RIS_INAC}

\end{document}